\documentclass[10pt,epsf]{article}
\usepackage{hyperref}
\usepackage{graphics}
\usepackage{amssymb}
\usepackage{epsf}
\usepackage{epsfig}

\def\ds#1{#1\kern-1ex\hbox{/}}
\def\dsh{h\kern-1.2ex /}

\newcommand{\bea}{\begin{eqnarray}}
\newcommand{\eea}{\end{eqnarray}}

\def\beq{\begin{equation}}
\def\eeq{\end{equation}}

\def\beqn{\begin{eqnarray}}
\def\eeqn{\end{eqnarray}}
\def\ba{\begin{eqnarray}}
\def\ea{\end{eqnarray}}

\newcommand{\beqa}{\begin{eqnarray}}
\newcommand{\eeqa}{\end{eqnarray}}

\begin{document}
\begin{center}
\vspace{1.cm}
{\bf \large Searching for an Axion-like Particle at the Large Hadron Collider\footnote{Presented by Marco Guzzi at IFAE 2009, Bari, 15-17 April 2009, Italy}\\ }
\vspace{1.5cm}
{\bf Claudio Corian\`{o}, Marco Guzzi, Antonio Mariano }

\vspace{1cm}

{\it $^a$Dipartimento di Fisica, Universit\`{a} del Salento \\
and  INFN Sezione di Lecce,  Via Arnesano 73100 Lecce, Italy}\\

\vspace{.5cm}

\begin{abstract}
Axion-like particles are an important part of the spectrum of anomalous gauge theories involving modified mechanisms of cancellation of the gauge anomalies. Among these are intersecting brane models, which are characterized by the presence of one physical axion. We overview  a recent study of their supersymmetric construction and some LHC studies of the productions rates for a gauged axion.

\end{abstract}
\end{center}
\newpage

\section{Introduction}
Particle physics candidates for dark matter abound, the most popular being the lightest neutralino of supersymmetric models such as the Minimal Supersymmetic Standard Model (MSSM) and its extensions. On the other hand, non-supersymmetric models also have their own dark matter candidate, the most popular one being the axion. Introduced by Peccei and Quinn \cite{Peccei:1977ur} long ago to solve the strong CP problem, the axion has required an extra global $U(1)$ symmetry, attached to the fields of the Standard Model (SM).

Its inclusion in a supersymmetric lagrangean has also always required the introduction of an extra global symmetry, together with a supersymmetric partner (the axino). To bring both components (the axion and the neutralino) under the spell of the gauge principle requires, quite likely, a gauging of the axionic symmetries.  

The study of axionic symmetries and of their gauging is an important aspect of string theory at lower energy, being connected with the presence, in these effective models, of several moduli fields, deprived of a potential and derived from geometrical compactifications. The attempt to match these descriptions with effective models (with or without supersymmetry and gravity) based on ordinary quantum field theory and supersymmetry (in flat space) is pursued in several works 
\cite{Coriano:2007fw,Coriano:2007xg} \cite{Anastasopoulos:2008jt}, where 
the role of the anomalous interactions and of light pseudoscalars is investigated under the tenets of gauge invariance and unitarity \cite{Armillis:2008bg}. 
In general, these effective field theories are characterized by the presence of higher dimensional operators to correct for the exposed (axial) anomaly (in the form of Wess-Zumino terms with St\"uckelberg axions) \cite{Coriano:2007fw}. Their gauge structure, in fact, requires at least one extra anomalous $U(1)$ interaction.
  
   The presence of gauge couplings and of a mass for this axion not related by the same suppression scale ($f_a$) - which is of the order of $10^{10}$ GeV for a traditional (ungauged) axion - makes this new axion a very attractive dark matter candidate. However, building axion models with gauged axionic symmetries is rather challenging from the field theory point of view since it requires a grasp of the unusual features of the chiral anomaly, from the organization of the effective action(s) to the presence of anomaly poles that challenge the consistency of the S-matrix  in some of their scattering amplitudes \cite{Coriano:2008pg}.
\section{Supersymmetric extensions}
One of the realizations of field theories which contain axion-like particles in their spectrum is the Minimal Low-Scale Orientifold Model (MLSOM) \cite{Coriano':2005js}, based on a construction involving charge assignments obtained from intersecting branes. These models introduce one St\"uckelberg axion for each anomalous $U(1)$ present in the gauge structure. In \cite{Coriano':2005js} it is shown that the physical spectrum of the MLSOM contains always one physical axion, independently of the number of anomalous U(1)'s. 

Supersymmetric extensions of these class of models have been discussed 
rather recently  \cite{Coriano:2008xa}, using a superpotential which is the one typical of the USSM \cite{Cvetic:1997ky}, extending previous studies Respect to the MSSM, the USSM contains an extra singlet superfield and an extra U(1) gauge symmetry, which is anomaly free. 
In the USSM-A model of  \cite{Coriano:2008xa}, which is the theory proposed as a possible supersymmetric extension of the MLSOM, this symmetry has been left anomalous, and the bosonic mechanism of cancellation of these anomalies, which is 
enforced via the inclusion of Wess-Zumino (counter)terms,  has been generalized. These models contain a combined Higgs-St\"uckelberg mechanism for the generation of the mass of the anomalous gauge bosons. 

A physical axion state (the axi-Higgs) appears quite naturally, together with its supersymmetric partner, the axino. 
This second state becomes a component of the fermionic neutralinos, after diagonalization of the mass matrix of the neutral fermion sector. 

Compared to other constructions, in which the axion disappeares from the physical spectrum being just a goldstone mode, in the USSM-A the presence of the extra singlet superfield allows a physical projection of the St\"uckelberg axion of the model on the axi-Higgs ($\chi$). Therefore this physical state inherits direct axion-like couplings (such as $\chi F\tilde{F}$) to the gauge fields, becoming a gauged supersymmetric axion \cite{Coriano:2008aw}. The particle is massless in the absence of Peccei-Quinn-breaking potentials, in which the axion appears as a phase, while the instanton vacuum can naturally lift its mass up to 
$10^{-4}$ eV as in the Peccei Quinn (invisible axion) case.

In the non-supersymmetric version of these theories, i.e. in the MLSOM, one of the special features of these theories is the presence of anomalous trilinear gauge interactions, which are totally absent in the Standard Model (SM) due to anomaly cancellation \cite{Armillis:2007tb}. We just recall that in the SM these interactions are suppressed and appear only away from the chiral limit in $AVV$ diagrams  (for instance in $Z \gamma \gamma$ vertices). We show in Fig. 1 results for the production of two axions $(\chi)$ in the MLSOM at the Tevatron (left panel) and at the LHC (right panel), as a function of the invariant mass $Q$ of the final state, obtained from a recent analysis. The gluon fusion chnnel is the most important one for their study. The processes involve the usual triangle diagrams as for the production of the Higgs sector. In our case we have considered two Higgs doublets $(h_0, H_0)$ in the CP-even sector. The s-channel exchange involves either an $h_0$, a $H_0$ or a $\chi$. 
We have selected a mass of the $h_0=15$ GeV and of $120$ GeV for the $H_0$, with $m_\chi=5$ GeV.

 \section{Conclusions}
 The search for light (gauged) pseudoscalars is for sure a challenge for the experimental acvitity at the LHC \cite{Armillis:2008vp}, requiring precise evaluation of the QCD background. However, while the detection of anomalous extra Z prime which modify the neutral current sector of these theories remains difficult, the identification of a light pseudoscalar (in the mass range of 1 to 10 GeV), which is their second signature, is more favoured.  
\begin{figure}[t]
\includegraphics[width=5cm, angle=-90]{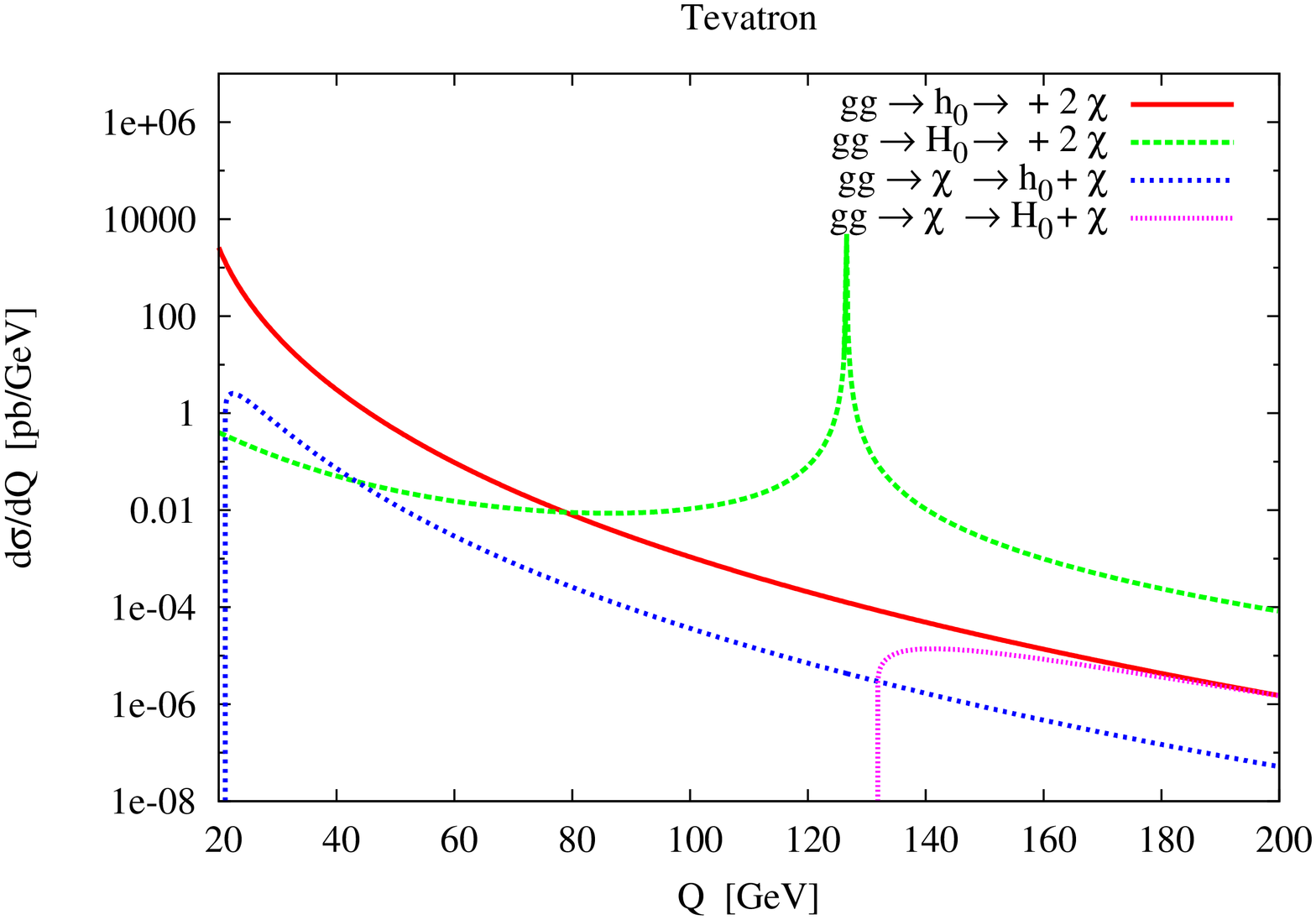}
\includegraphics[width=5cm, angle=-90]{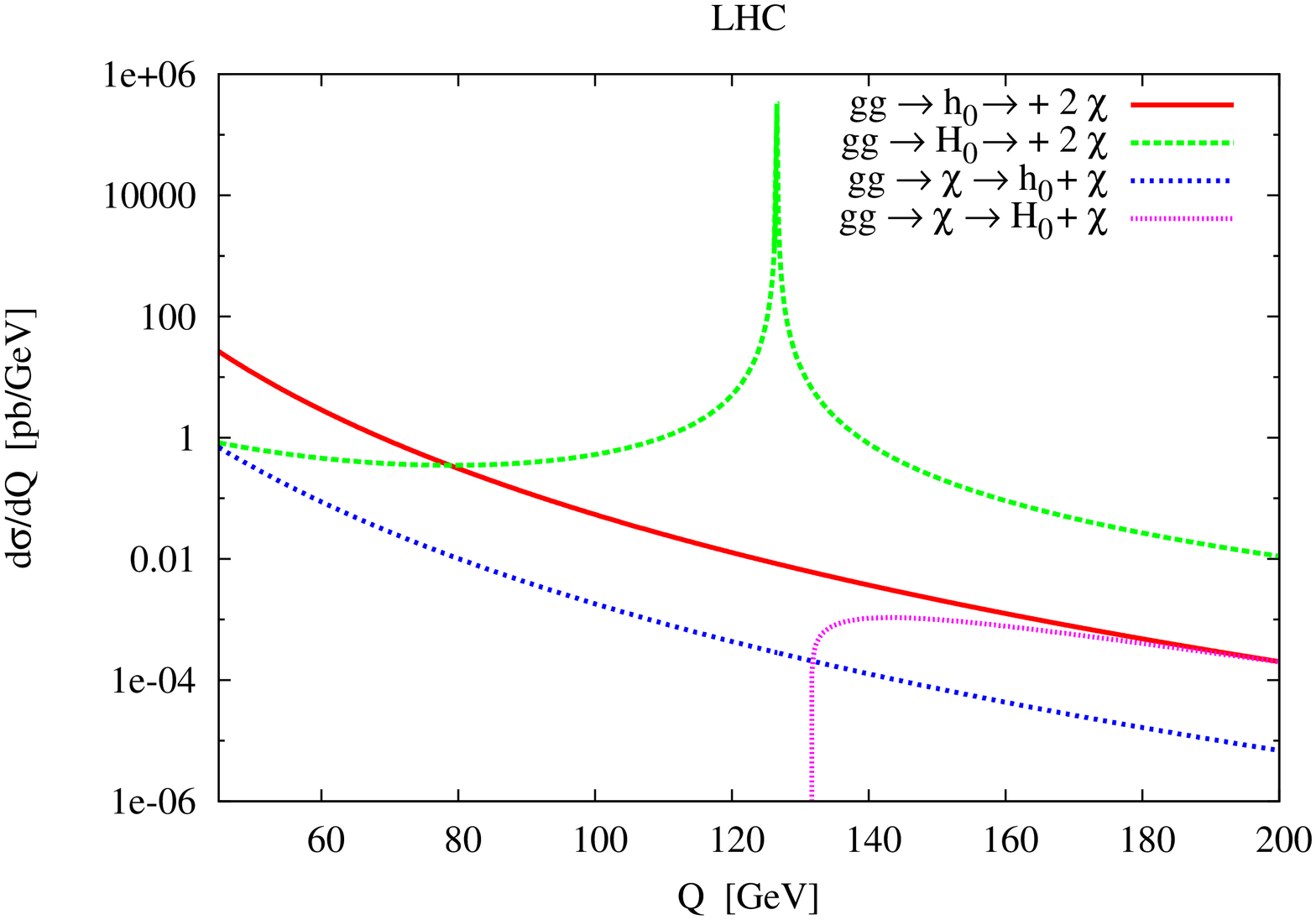}
\caption{\small $gg\rightarrow$ 2-scalar processes mediated by a three-linear vertex at the Tevatron and at the LHC.}
\label{2scalar_prod_LHC}
\end{figure}
\newpage
\centerline {\bf Acknowledgments}
We thank Roberta Armillis and Luigi Delle Rose for discussions.
This work was supported (in part) by the European Union through the Marie Curie Research and Training Network ``Universenet'' (MRTN-CT-2006-035863).


\begin{thebibliography}{9}

\bibitem{Peccei:1977ur}
R.D. Peccei and H.R. Quinn,
\newblock Phys. Rev. D16 (1977) 1791.

\bibitem{Coriano:2007fw}
C. Corian\`o, N. Irges and S. Morelli,
\newblock JHEP 07 (2007) 008, hep-ph/0701010.

\bibitem{Coriano:2007xg}
C. Corian\`o, N. Irges and S. Morelli,
\newblock Nucl. Phys. B789 (2008) 133, hep-ph/0703127.
\bibitem{Anastasopoulos:2008jt} 
P. Anastasopoulos, F. Fucito, A. Lionetto, G. Pradisi, A. Racioppi, Y. S. Stanev
\newblock Phys.Rev.D78:085014, 2008. 


\bibitem{Armillis:2008bg}
R. Armillis et~al.,
\newblock JHEP 10 (2008) 034, 0808.1882.

\bibitem{Coriano:2008pg}
C. Corian\`o, M. Guzzi and S. Morelli,
\newblock Eur. Phys. J. C55 (2008) 629, 0801.2949.

\bibitem{Coriano':2005js}
C. Corian\`o, N. Irges and E. Kiritsis,
\newblock Nucl. Phys. B746 (2006) 77, hep-ph/0510332.

\bibitem{Coriano:2008xa}
C. Corian\`o, M. Guzzi, A. Mariano, S. Morelli,
\newblock (2008), 0811.3675.

\bibitem{Cvetic:1997ky}
M. Cvetic, D.A. Demir, J.R. Espinosa , L.L Everett, P. Langacker, 
\newblock Phys. Rev. D56 (1997) 2861, hep-ph/9703317.

\bibitem{Coriano:2008aw}
C. Corian\`o, M. Guzzi, N. Irges, Nikos,  A. Mariano,
\newblock (2008), 0811.0117.

\bibitem{Armillis:2007tb}
R. Armillis, C. Corian\`o and M. Guzzi,
\newblock JHEP 05 (2008) 015, 0711.3424.

\bibitem{Armillis:2008vp}
R. Armillis, C. Corian\`o, M. Guzzi, S. Morelli,
\newblock (2008), 0809.3772.

\end{thebibliography}
\end{document}